\documentclass[useAMS,usenatbib]{mn2e}
\usepackage{natbib}
\usepackage{aas_macros}
\usepackage{amssymb}
\usepackage{psfig}
\usepackage{graphicx}
\usepackage{dcolumn}
\usepackage{bm}
\usepackage{multirow}

\title[Neutral Hydrogen Tully Fisher Relation]{Neutral Hydrogen Tully Fisher Relation:\\ The case for Newtonian Gravity}

\author[Chakraborti \& Khedekar]{
  \parbox[t]{16cm}{
Sayan Chakraborti$^{1}$\thanks{sayan@tifr.res.in},
Satej Khedekar$^{1}$
  }\\
$^{1}$Department of Astronomy and Astrophysics, Tata Institute of Fundamental Research, Mumbai 400 005, India\\
}
\begin{document}

\date{Draft \today}

\pagerange{\pageref{firstpage}--\pageref{lastpage}} \pubyear{2011}

\maketitle

\label{firstpage}

\begin{abstract}
Intrinsic luminosities are related to rotation
velocities of disk galaxies by Tully Fisher (TF) relations.
The Baryonic TF (BTF) relation has recently been explained with Dark Matter
and Newtonian Gravity as well as Modified Newtonian Dynamics (MOND).
However, recent work has pointed out that the currently used
BTF relation ignores the contribution from hot gas and 
oversimplifies complex galaxy-scale physics.
In this Letter,
we advocate the use of the Neutral Hydrogen TF (HITF) relationship, which
is free from dust obscuration and stellar evolution effects, as a clean
probe of gravity and dynamics in the weak field regime.
We incorporate the physics of hot gas from supernova feedback which
drives the porosity of the Interstellar Medium (ISM). A simple model
that includes supernovae feedback, is generalized to include a
parametrized effective gravitational force law. We test our model against
a catalogue of galaxies, spanning the full range of disks from dwarf galaxies
to giant spirals, to demonstrate that a Kennicutt-Schmidt (KS) law for star
formation and simple Newtonian gravity is adequate for explaining the
observed H{\sc i} scaling relations. The data rules out MOND-like theories, within
the scope of this model.
\end{abstract}


\section{Introduction}
If all matter is accounted for,
Newtonian gravity and dynamics
should provide an adequate description of the motion of celestial
bodies in the weak gravity regime. Yet, \citet{1937ApJ....86..217Z}
found that the luminous matter in galaxies (then called Nebulae)
could not account for the dynamics in galaxy clusters. This
{\it Dark Matter} (DM) problem was extended to galaxies with the
advent of resolved rotation curves \citep{1970ApJ...159..379R}.
\citet{1983ApJ...270..365M} proposed a Modified Newtonian
Dynamics (MOND), as an alternative to the DM paradigm.
Galactic dynamics provide us a way to test the nature of
gravity in the weak field limit and settle one of the most
important outstanding questions in physics.

\citet{1977A&A....54..661T} (TF) found a correlation between the global
neutral hydrogen (H{\sc i}) profile widths and absolute optical magnitudes of
disk galaxies.
This relationship has since then, been well calibrated
against Cepheid distances \citep{2000ApJ...529..698S} and widely used
as a distance indicator \citep{2000ApJ...529..698S}.
Because the absolute magnitudes are essentially
a mass proxy and the profile widths are a velocity proxy, this relation
implies a connection between the mass and dynamics.
\citet{2000ApJ...533L..99M} have shown that even though gas dominated
galaxies with a low stellar fraction deviate from the TF relation, they
fall on a ``more fundamental'' relation between the total baryonic mass
and the rotation velocity.
\citet{2011PhRvL.106l1303M} has shown that
the BTF relation in gas rich galaxies is consistent with the prediction
from MOND.
\citet{2011PhRvL.106l1303M} has included only gas rich galaxies in the
analysis, to minimize the uncertainty in the baryonic mass
introduced by the error in stellar masses.
However the selection by gas fraction restricts the range of
galaxy masses and rotation velocities. This may also bias the slope
as the gas fraction is a function of the total mass.
\citet{2010NatPh...6...96L} has also explained the BTF, but within the framework
of Newtonian gravity.
Hence, the BTF can plausibly be explained in both
Newtonian and MOND gravities, using a suitably selected model.
Any such test is therefore intrinsically model dependent.

An attempt to explain a TF relation must also take into account,
instabilities in the disk \citep{1997ApJ...482..659D}, resulting star
formation and supernova feedback \citep{1997ApJ...481..703S}.
\citet{2011arXiv1108.2271G} has pointed out that BTF only accounts
for stars, molecular, and atomic gas, while the contribution of
ionized gas is almost universally missed. Such a situation has
led \citet{2011arXiv1108.5734F} to comment that the claimed Cold DM
prediction for the BTF, by \citet{2011PhRvL.106l1303M},
is a gross oversimplification of the complex galaxy-scale physics involved.
Our work, for the first time, accounts for the role of supernova driven hot
gas in determining the H{\sc i} scaling relations.
In fact 
\citet{1997ApJ...481..703S} has explained the TF relation in the
B-band, which traces recent star formation, by considering self regulated
star formation. However, the optical photons are strongly affected
by dust obscuration and inclination,
which may significantly bias the TF relations. Using Near Infrared
TF relations \citep{1997AJ....113...53G,2006ApJ...653..861M}
can minimize dust effects and provide tighter correlations. 
The alternative is to use a relation between H{\sc i} masses
and H{\sc i} profile widths, which we henceforth refer to as the H{\sc i} Tully
Fisher (HITF) relation. This relation is far less sensitive to obscuration
as the universe is largely transparent at radio
wavelengths \citep{2011PhRvL.106v1301K}. The H{\sc i} signal originates
from the ISM and is therefore independent of the
stellar ages in the galaxy.

Any model which explains this relation must also
explain \citep{2011ApJ...732..105C} the observed \citep{2003ApJ...585..256R,
2008MNRAS.386.1667B} scaling relation between H{\sc i} masses and areas of disk
galaxies. In this Letter we further develop an analytic model by
\citet{1997ApJ...481..703S} and \citet{2011ApJ...732..105C} to derive the
H{\sc i} masses of disk galaxies, taking into account the competition between
gravitational instabilities and mechanical feedback from supernovae. We also
generalize the model to include arbitrary $(\sqrt{M}/r)^\gamma$ laws of
gravity and test it against our catalogue of galaxies. The HITF data spans
the full range of disks from dwarfs galaxies observed with the
Giant Metrewave Radio Telescope
(GMRT) \citep{2008MNRAS.386.1667B}, to giant spirals observed with the
Arecibo \citep{2010AJ....140..663G}. The H{\sc i} mass vs area data is from
\citet{2003ApJ...585..256R}. We conduct a joint likelihood analysis of 
these data sets within the scope of our model to demonstrate the
validity of Newtonian gravity at galactic scales, and also rule out
MOND.

\section{H{\sc i} Scaling Relations}
H{\sc i} scaling relations in galaxies have been known for the past
two decades. \citet{1993ApJ...417..494B} had already found a weak scaling
($\Delta V \propto M_{\rm HI}^{\sim 0.3}$) of the velocity width ($\Delta V$)
with the H{\sc i} mass ($M_{\rm HI}$). This however implied a strong
dependence of $M_{\rm HI}$ on $\Delta V$ with a power law index between
2.1 and 3.6 \citep{1993ApJ...417..494B}. The optical TF relation,
which measures stellar mass, breaks down at the low luminosity end
\citep{2000ApJ...533L..99M}. However no such breaks have been reported in the
H{\sc i} scaling. We are now in a position to extend the HITF to the full
range of galaxy masses for the first time and show in this work that it is
consistent with a single scaling relation.

Scaling relations between $M_{\rm HI}$ and the surface area's of galaxies
also have a long history. \citet{1983AJ.....88..881G} reported that
H{\sc i} sizes and masses were found to be correlated in the same
manner, irrespective of whether they were H{\sc i} rich or H{\sc i} poor.
Optical sizes and H{\sc i} masses were also seen to be correlated
\citep{1984AJ.....89..758H,2001A&A...370..765V}.
\citet{2003ApJ...585..256R} reported that
$M_{\rm HI}$ and surface areas of ADBS \citep{2000ApJS..130..177R}
galaxies are consistent with a nearly constant average H{\sc i} surface density
of the order $\sim10^7M_\odot$ kpc$^{-2}$. \citet{2011ApJ...732..105C}
has recently explained this surprising correlation, 
as the result of self-regulated star formation, driven by the
competition between gravitational instabilities in a rotationally
supported disk and mechanical feedback from supernovae.

\section{Generalized Model}
\label{model}
Following \citet{1997ApJ...481..703S}, we consider a rotationally
supported gas disk, fragmenting and forming stars due to gravitational
instability. In such a scenario the asymptotic velocity can be
related to the enclosed dynamical mass ($M_{\rm d}$). However this
requires the knowledge
of an effective force law for gravity in the weak field limit.
Since we wish to determine the nature of gravity at
these scales, it is necessary not to use a particular variant, but
a parametrization which can encompass a wide range of theories including
Newtonian gravity and MOND\footnote{MOND can be thought of as a modification
of either the inertial term ($F=\mu ma$) or of the
effective gravitational force term ($F=F_{\rm Newton} / \mu$)
by $\mu (a/a_0)$, where $a_0$ is the acceleration scale below which the
dynamics is modified. At the
low acceleration regime, relevant for disk galaxies, they result in the
same rotation curves \citep{2002ARA&A..40..263S}.}. We therefore, parametrize
 the gravitational force ($F$) at a radius $r$ as,
\begin{equation}
 F \propto \left( \frac{\sqrt{M_{\rm d}}}{r} \right)^{\gamma} ,
\end{equation}
where $\gamma$ is a free parameter which is to be determined from observations.
Here, $\gamma=2$ gives back the simple Newtonian case, while $\gamma=1$
gives the familiar MOND case in the low acceleration regime relevant
at galactic scales.

To relate the gravitational force to a rotation velocity, one needs a
prescription for $M_{\rm d}$. \citet{2009MNRAS.394..340G} points out that
the observed dynamical mass scales as cube of the optical radius.
\citet{2011PhRvL.106q1302L} have shown that dark matter particles
interacting through a Yukawa potential could provide a natural explanation
for a characteristic density in dark matter dominated halos.
We exploit this observed \citep{2009MNRAS.394..340G}
relation
$M_{\rm d} \propto R^3$,
where $R$ is the size of the H{\sc i} disk,
to express the rotation velocity as,
\begin{equation}
 V \propto R^{(2+\gamma)/4} .
\label{vr}
\end{equation}
Thus the dependence of the asymptotic rotation velocity
on disk size is a function of $\gamma$ and therefore is a probe of
the effective gravitational force law in the low acceleration case.


\begin{figure}
 \centering
  \includegraphics[width=\columnwidth]{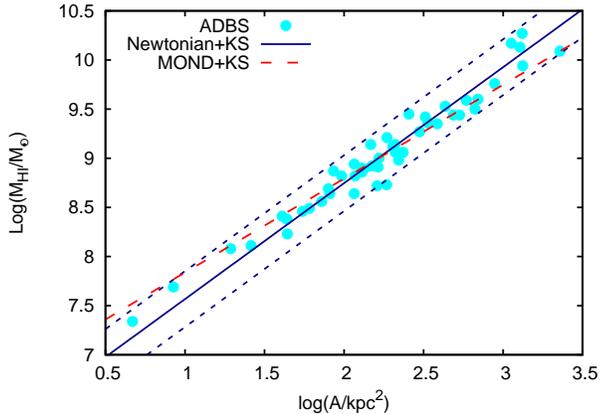}
  \caption{$M_{\rm HI}$ vs $A$ data for ADBS galaxies (cyan balls)
  from \citet{2003ApJ...585..256R}. Statistical uncertainties are
  comparable to the
  sizes of the balls. Blue solid (and dotted) line shows the
  Newtonian$+$KS model (and 2$\sigma$ scatter region) with $\gamma=2$
  and $\beta=1.4$ put in Eqn.~\ref{eq:MA}. Red dashed line shows
  the MOND$+$KS model with $\gamma=1$ and $\beta=1.4$.}
  \label{fig:mvsa}
\end{figure}

\citet{2011ApJ...732..105C} has argued that the observed correlation between H{\sc i}
mass and surface area may be the result of self regulated star formation. Even
when a disk galaxy is driven away from this fundamental line, self regulation
of the porosity (driven by supernova feedback \citep{1979ARA&A..17..213M}
of the ISM and gravitational instability \citep{1960AnAp...23..979S,1964ApJ...139.1217T}
 of the star forming disk,
can control the star formation rate. This ensures that the star forming
disk is always marginally unstable. Under these considerations
\citet{1997ApJ...481..703S} has shown that the star formation rate
(SFR) scales as,
${\rm SFR} \propto V^{5/2}$,
where $V$ is the asymptotic rotation velocity of the disk.

We can now express the surface density of the SFR ($\Sigma_{\rm SFR}$) as
$\Sigma_{\rm SFR}\equiv {\rm SFR}/({\pi R^2})\propto V^{5/2}R^{-2} $.
Substituting for the rotation velocity ($V$) as a function of $R$
from Eqn.~\ref{vr}, we have
\begin{equation}
 \Sigma_{\rm SFR} \propto R^{(5(2+\gamma)/8) -2}
                  \propto R^{(5\gamma-6)/8}.
\end{equation}
This relates the surface density of SFR to the size of the galaxy.
However, the $\Sigma_{\rm SFR}$ is also related to the surface density
of cold gas according to the KS law \citep{1998ARA&A..36..189K}.
If the cold gas in a galaxy is dominated by H{\sc i}, then we can use the KS law
to express the SFR surface density as
$\Sigma_{\rm SFR} \propto \Sigma_{\rm HI}^{\beta}$,
where \citet{1998ARA&A..36..189K} gives the power law index
$\beta$ a value of $1.4 \pm 0.15$.

Eliminating $\Sigma_{\rm SFR}$, we
get the H{\sc i} surface density as
$\Sigma_{\rm HI}^\beta \propto R^{(5\gamma-6)/8}$.
Integrating the surface density over the size of the galaxy, we get
\begin{equation}
 M_{\rm HI} \equiv \pi R^2 \Sigma_{\rm HI} \propto R^{((5\gamma-6)/(8\beta))+2} .
\end{equation}
Expressing this as a function of the surface area, we get
\begin{equation}
 M_{\rm HI} = C_{\rm A} A^{((5\gamma-6)/(16\beta))+1},
\label{eq:MA}
\end{equation}
where $C_{\rm A}$ is a constant of proportionality.
This generalizes the result ($M_{\rm HI} \propto A^{1.18}$ for
Newtonian gravity and KS law slope of 1.4) from
\citet{2011ApJ...732..105C} to arbitrary
KS law slopes and generalized effective gravitational force laws.
\citet{2011PhRvL.106v1301K} have shown that such a relation depends only
weakly on the metallicity of the galaxy.
The power law relates the H{\sc i} mass of a galaxy to its size and can be directly
compared to the data (See Fig. \ref{fig:mvsa}).

We can now use the scaling in Eqn.~\ref{vr} between $V$ and $R$ to find $M_{\rm HI}$ as
a function of the galaxy size. This gives us the H{\sc i} mass as
\begin{equation}
 M_{\rm HI} = C_{\rm V} V^{(5\gamma + 16\beta-6)/(2\beta (2 + \gamma))},
\label{eq:MV}
\end{equation}
where $C_{\rm V}$ is another constant of proportionality.
This shows that the observed slope of the HITF relation is a probe
of both the KS law and the law of gravity. For Newtonian gravity and
KS law slope of 1.4 we have $M_{\rm HI} \propto V^{2.36}$.
This can now be compared directly
with the data (See Fig. \ref{fig:hitf}),
but a prior knowledge of the KS law slope will be required
to determine the effective gravitational force law. This shortcoming can be
overcome by simultaneously comparing Eqn.~\ref{eq:MV} to the HITF data and
Eqn.~\ref{eq:MA} to the H{\sc i} mass vs area data.
The ratio of the two slopes is a function of $\gamma$, but independent of
$\beta$. Hence, determining both slopes from observations can
break the $\gamma-\beta$ degeneracy and determine $\gamma$ irrespective
of our prior knowledge of $\beta$. Being able to explain
both observed relations for plausible values of these parameters
is a crucial reality check for our model.

\section{Catalogue of Galaxies}
The Arecibo Dual-Beam Survey \citep{2000ApJS..130..177R} (ADBS) has
conducted a ``blind'' survey of $\sim$ 430 deg$^2$ of sky and detected the 
H{\sc i} signal in 265 galaxies. Interferometric mapping of 84 galaxies was carried
out with the NRAO's Very Large Array (VLA) as most of the ADBS galaxies
were unresolved at the
resolution of the Arecibo. Accurate sizes of 50 of them were compiled
by \citet{2003ApJ...585..256R} to tabulate $M_{\rm HI}$ and $A$
(H{\sc i} cross section with column density above $2\times10^{20} \ {\rm cm^{-2}}$)
of individual ADBS galaxies. The observed sizes span 3 orders of
magnitude ($0.5>{\rm log}(A/{\rm kpc^2})>3.5$)
and provide the data for the H{\sc i} mass vs area relation.
For the following analysis, we have used only objects with large projected
areas (${\rm log}(A/{\rm kpc^2})>2$) as the smaller galaxies may have
systematically uncertain sizes limited by the resolution of the present surveys.
Fig. \ref{fig:mvsa} presents the observed data and compares it with the
fiducial models.


\begin{figure}
 \centering
  \includegraphics[width=\columnwidth]{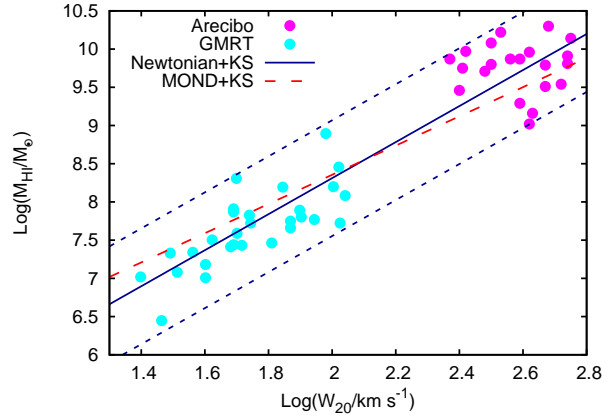}
  \caption{$M_{\rm HI}$ vs inclination corrected $W_{20}$ data for
  galaxies from Arecibo
  (magenta balls) and GMRT (cyan balls). See text for sources of data.
  Statistical uncertainties are comparable to the sizes of the balls.
  Blue solid
  (and dotted) line shows the Newtonian$+$KS model (and 2$\sigma$ scatter
  region) with $\gamma=2$ and $\beta=1.4$ put in Eqn.~\ref{eq:MV}.
  Red dashed line shows the MOND$+$KS model with $\gamma=1$ and
  $\beta=1.4$.}
  \label{fig:hitf}
\end{figure}

\citet{2010AJ....140..663G} presented a sample of local ($D<60$ Mpc)
field galaxies with accurate 21 cm observations to construct a BTF
relation. We select the galaxies listed by \citet{2000ApJ...529..698S}
and used by \citet{2010AJ....140..663G}. All these galaxies have
accurate Cepheid distances \citep{2000ApJ...529..698S} determined using
the {\it Hubble Space Telescope} $H_0$ Key Project and 21 cm Arecibo
observations from \citet{1997AJ....113...53G}. For all these galaxies
we use $M_{\rm HI}$ and $W_{20}$ (the width at which the H{\sc i} profile
drops to 20\% of its maximum value) as listed by \citet{2010AJ....140..663G},
inclination corrected following the method of \citet{1997AJ....113...53G}.
Following \citet{1977A&A....54..661T} this is commonly used as a proxy
for rotation velocity in radio astronomy \footnote{It is possible
to infer a rotation velocity $V_{\rm rot}$ by fitting, say a tilted
ring model (with a poorly determined inclination) to the H{\sc i} data cube,
but this is model dependent.
\citet{2008MNRAS.386..138B} confirm that $W_{20}$ correlates better with
mass indicators than $V_{\rm rot}$. In the rest of the analysis we
replace $V_{\rm rot}$ with its observational proxy $W_{20}$.}.
This provides the data for the high mass range in the HITF relation.

To accurately determine the slope of the HITF relation, this catalogue needs
to be supplemented with data on low mass galaxies as well. The
Faint Irregular Galaxies GMRT Survey (FIGGS) conducted by 
\citet{2008MNRAS.386.1667B} aims at characterizing the neutral
ISM properties of faint, gas-rich dwarf galaxies
through 21 cm GMRT observations. This provides us with the $M_{\rm HI}$
and $W_{20}$ (again inclination corrected following Ref.~\citep{1997AJ....113...53G})
measurements at the low mass end of the HITF relation.
At the low velocity end, line broadening due to velocity dispersion from
turbulent motion in the H{\sc i}, may play a significant role. We correct
for this effect following the prescription from \citet{1985ApJS...58...67T}
\footnote{This may introduce a systematic uncertainty. However, according to
Eqn.~5 of \citet{2005MNRAS.364.1337S}, a $10\%$ variation in the gas
density, leads to a $\sim1\%$ change in the gas dispersion velocity. This
will introduce a $\lesssim2\%$ systematic error in the corrected velocity
width, which is small compared to the statistical uncertainty of $\sim10\%$.}.
Fig \ref{fig:hitf} displays the observational HITF data and compares it with the
fiducial models.

\begin{figure}
 \centering
  \includegraphics[width=\columnwidth]{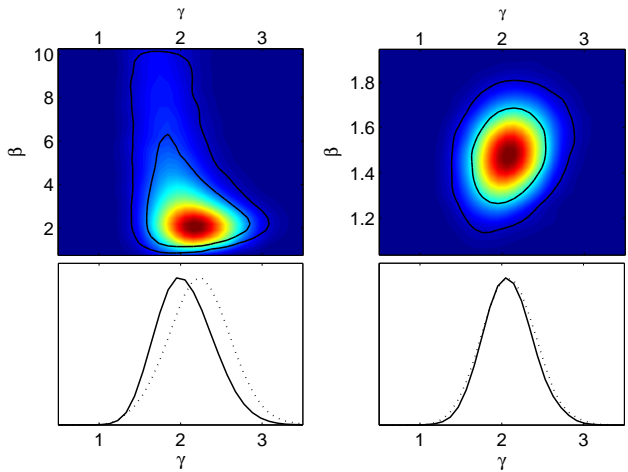}
  \caption{The hotter (cooler) colours denote regions with higher (lower)
 mean likelihoods. The solid (dotted) lines denote the 1$\sigma$/2$\sigma$
 confidence limits of the marginalized (mean) likelihoods.
 The figures on the right include a Gaussian prior on $\beta$, 
 while the figures on the left
 do not include such an external prior. Note that the value of $\gamma$=1, 
 corresponding to the case of the
 MOND gravity, is completely ruled out, while the Newtonian gravity ($\gamma$=2)
 gives an excellent
 fit to the data.}
  \label{fig:likelihood}
\end{figure}

\section{MCMC estimation}
In order to obtain an effective law of gravity at the galactic scales
using the data on
$M_{\rm HI}$, $A$ and $v$, we consider the two scaling relations -- $M_{\rm HI}$
 vs $A$ and $M_{\rm HI}$ vs $v$, as expressed through Eqn.~\ref{eq:MA} and
 \ref{eq:MV} respectively. We compute $\chi^2$ for the the data
as a sum, $\chi^2 = \chi_A^2 +  \chi_v^2 $, one
for each for the relations -- $M_{\rm HI}$ vs A and $M_{\rm HI}$ vs $v$,
respectively. For each $\chi^2$. The weight, $\sigma$, is computed by
adding the intrinsic scatter and observed error in quadrature.

We then compute the likelihoods, $\exp(-\chi^2 /2)$ in the four
 dimensional parameter space of $\{C_{\rm A}, \ C_{\rm V}, \ \beta, \
\gamma \}$.
We perform Markov Chain Monte Carlo (MCMC) runs\footnote{An MCMC
algorithm is useful
for sampling the likelihoods from a high dimensional probability
distribution based on constructing
Markov chains that have the desired distribution as their equilibrium
distribution,
 {\it i.e.} $\exp(-\chi^2 /2)$; see \cite{2002PhRvD..66j3511L} for more details.} in the
using these likelihoods. 
Of these parameters, we eventually marginalize over the nuisance
parameters $C_{\rm A}$ and $C_{\rm V}$.
Note that we do not assume a completely deterministic
form for the KS law, but require only a simple power law relation, which is
eventually marginalized when quoting our final results on $\gamma$.

The results of our likelihood analysis is shown in Fig.
\ref{fig:likelihood}. The top figures
displays the 1$\sigma$ and 2$\sigma$ constraints
in the $\beta-\gamma$ plane. The top left figure shows the likelihoods without
any external prior on $\beta$, while the top right figure shows the results
after inclusion of a Gaussian prior of $\beta=1.4\pm0.15$ as obtained
by \citet{1998ARA&A..36..189K} through independent observations.
The corresponding figures at the bottom show the marginalized (over $\beta$)
1-D distribution on $\gamma$.

Our MCMC exploration of the parameter space, without the priors on
the star formation law, reveals that even though $\beta=2.01_{-1.26}^{+4.23}$
is not well constrained, the effective law of gravity is highly constrained
with $\gamma$=2.06$\pm0.34$, consistent with Newtonian gravity ($\gamma=2$)
but clearly ruling out MOND ($\gamma=1$).
At the same time, the best fit region in the $\beta-\gamma$ plane
also includes the region with KS law slope of $\beta$=1.4$\pm 0.15$ from
\citet{1998ARA&A..36..189K}. Hence our result confirms that the observed
HI scaling relation is not only consistent with Newtonian gravity but also
with the KS law prescription for star formation
\footnote{\citet{2011MNRAS.414L..55R} have found that when comparing
$\Sigma_{\rm HI}$ to the ${\rm H\alpha}$ emission, the KS law is steeper
($\beta\sim2$), which is also consistent with our result.
A summary of the impact of different star formation laws on galaxy
formation is given by \citet{2010arXiv1011.5506L}.}.
We find that both with and without an external prior on $\beta$,
our model can conclusively rule out the MOND model of gravity; the constraints
on $\gamma$ get slightly better after including the prior on $\beta$.
This shows that our results are insensitive to the prior used on the KS law,
as expected from Section \ref{model}.
The likelihood analysis yields values of $\gamma$=$2.08\pm0.29$ after
including an external prior on $\beta$.

\section{Discussions}
TF relations are being increasingly discussed (and also criticised)
in the context of weak field tests of gravity. In this letter we have
suggested the simultaneous use of two H{\sc i} scaling relations
for this purpose. Our test is free from some of the systematics and
selection bias that affect the existing methods. Our model
incorporates, for the first time, the physics of supernova feedback
driven hot gas into this analysis.
We have shown that a simple model of self regulated star formation
accounts for the observed H{\sc i} scaling relations in disk galaxies.

The crucial ingredients which make the model agree with the data are the star
formation law and the effective law of gravity in the weak field limit.
Earlier work shows that rotation curves and single scaling relationships,
such as the BTF, can be explained both within the DM \citep{2010NatPh...6...96L}
and MOND \citep{2011PhRvL.106l1303M} paradigms.
Hence, a new method was required to use galactic dynamics for a weak
field test of gravity. Our work shows that the HITF and the M-vs-A
relations taken together can constrain the effective law of gravity
relevant on the galactic scale.
The data rules out MOND at more than $99.9\%$ confidence at galactic
scales within the scope of this model.
We show that our model suffices to explain the data using a simple KS law
prescription for star formation and Newtonian gravity. This is a triumph
for the DM paradigm with Newtonian gravity in the low acceleration regime. 

With Extended-VLA follow up of ALFALFA \citep{2005AJ....130.2598G}
detected galaxies and the
the advent of high sensitivity, high resolution, next generation
radio telescopes such as the Square Kilometer Array, the number of H{\sc i}
detected galaxies is set to grow rapidly \citep{2011PhRvL.106v1301K}.
Our study implies that the
soon to be observed scaling relations between H{\sc i} masses of these
galaxies, their rotation velocities and sizes will provide an
excellent probe for the nature of gravity and dynamics in the weak
field limit.

\section*{Acknowledgements}
The authors would like to thank Alak Ray, Subhabrata Majumdar, Biman Nath
for discussions
and an anonymous referee for comments. The NRAO is a facility of the NSF
operated under cooperative agreement by Associated Universities, Inc.
GMRT is run by the National Centre for Radio Astrophysics of the Tata
Institute of Fundamental Research.

\bibliographystyle{mn2e}
\bibliography{hitf}

\begin{thebibliography}{}

\bibitem[\protect\citeauthoryear{{Begum} et~al.,}{{Begum}
  et~al.}{2008a}]{2008MNRAS.386..138B}
{Begum} A.,  et~al., 2008a, \mnras, 386, 138

\bibitem[\protect\citeauthoryear{{Begum} et~al.,}{{Begum}
  et~al.}{2008b}]{2008MNRAS.386.1667B}
{Begum} A.,  et~al., 2008b, \mnras, 386, 1667

\bibitem[\protect\citeauthoryear{{Briggs} \& {Rao}}{{Briggs} \&
  {Rao}}{1993}]{1993ApJ...417..494B}
{Briggs} F.~H.,  {Rao} S.,  1993, \apj, 417, 494

\bibitem[\protect\citeauthoryear{{Chakraborti}}{{Chakraborti}}{2011}]{2011ApJ.%
..732..105C}
{Chakraborti} S.,  2011, \apj, 732, 105

\bibitem[\protect\citeauthoryear{{Dalcanton}, {Spergel} \&
  {Summers}}{{Dalcanton} et~al.}{1997}]{1997ApJ...482..659D}
{Dalcanton} J.~J.,  {Spergel} D.~N.,    {Summers} F.~J.,  1997, \apj, 482, 659

\bibitem[\protect\citeauthoryear{{Foreman} \& {Scott}}{{Foreman} \&
  {Scott}}{2011}]{2011arXiv1108.5734F}
{Foreman} S.,  {Scott} D.,  2011, ArXiv e-prints

\bibitem[\protect\citeauthoryear{{Garcia-Appadoo} et~al.,}{{Garcia-Appadoo}
  et~al.}{2009}]{2009MNRAS.394..340G}
{Garcia-Appadoo} D.~A.,  et~al., 2009, \mnras, 394, 340

\bibitem[\protect\citeauthoryear{{Giovanelli} et~al.,}{{Giovanelli}
  et~al.}{1997}]{1997AJ....113...53G}
{Giovanelli} R.,  et~al., 1997, \aj, 113, 53

\bibitem[\protect\citeauthoryear{{Giovanelli} et~al.,}{{Giovanelli}
  et~al.}{2005}]{2005AJ....130.2598G}
{Giovanelli} R.,  et~al., 2005, \aj, 130, 2598

\bibitem[\protect\citeauthoryear{{Giovanelli} \& {Haynes}}{{Giovanelli} \&
  {Haynes}}{1983}]{1983AJ.....88..881G}
{Giovanelli} R.,  {Haynes} M.~P.,  1983, \aj, 88, 881

\bibitem[\protect\citeauthoryear{{Gnedin}}{{Gnedin}}{2011}]{2011arXiv1108.2271%
G}
{Gnedin} N.~Y.,  2011, ArXiv e-prints

\bibitem[\protect\citeauthoryear{{Gurovich} et~al.,}{{Gurovich}
  et~al.}{2010}]{2010AJ....140..663G}
{Gurovich} S.,  et~al., 2010, \aj, 140, 663

\bibitem[\protect\citeauthoryear{{Haynes} \& {Giovanelli}}{{Haynes} \&
  {Giovanelli}}{1984}]{1984AJ.....89..758H}
{Haynes} M.~P.,  {Giovanelli} R.,  1984, \aj, 89, 758

\bibitem[\protect\citeauthoryear{{Kennicutt}
  Jr.}{{Kennicutt}}{1998}]{1998ARA&A..36..189K}
{Kennicutt} Jr. R.~C.,  1998, \araa, 36, 189

\bibitem[\protect\citeauthoryear{{Khedekar} \& {Chakraborti}}{{Khedekar} \&
  {Chakraborti}}{2011}]{2011PhRvL.106v1301K}
{Khedekar} S.,  {Chakraborti} S.,  2011, Phys. Rev. Lett., 106, 221301

\bibitem[\protect\citeauthoryear{{Lagos} et~al.,}{{Lagos}
  et~al.}{2010}]{2010arXiv1011.5506L}
{Lagos} C.~d.~P.,  et~al., 2010, arXiv:1011.5506

\bibitem[\protect\citeauthoryear{{Larson}}{{Larson}}{2010}]{2010NatPh...6...96%
L}
{Larson} R.~B.,  2010, Nature Physics, 6, 96

\bibitem[\protect\citeauthoryear{{Lewis} \& {Bridle}}{{Lewis} \&
  {Bridle}}{2002}]{2002PhRvD..66j3511L}
{Lewis} A.,  {Bridle} S.,  2002, \prd, 66, 103511

\bibitem[\protect\citeauthoryear{{Loeb} \& {Weiner}}{{Loeb} \&
  {Weiner}}{2011}]{2011PhRvL.106q1302L}
{Loeb} A.,  {Weiner} N.,  2011, Phys. Rev. Lett., 106, 171302

\bibitem[\protect\citeauthoryear{{Masters}, {Springob}, {Haynes} \&
  {Giovanelli}}{{Masters} et~al.}{2006}]{2006ApJ...653..861M}
{Masters} K.~L.,  {Springob} C.~M.,  {Haynes} M.~P.,    {Giovanelli} R.,  2006,
  \apj, 653, 861

\bibitem[\protect\citeauthoryear{{McCray} \& {Snow} Jr.}{{McCray} \&
  {Snow}}{1979}]{1979ARA&A..17..213M}
{McCray} R.,  {Snow} Jr. T.~P.,  1979, \araa, 17, 213

\bibitem[\protect\citeauthoryear{{McGaugh}}{{McGaugh}}{2011}]{2011PhRvL.106l13%
03M}
{McGaugh} S.~S.,  2011, Phys. Rev. Lett., 106, 121303

\bibitem[\protect\citeauthoryear{{McGaugh} et~al.,}{{McGaugh}
  et~al.}{2000}]{2000ApJ...533L..99M}
{McGaugh} S.~S.,  et~al., 2000, \apjl, 533, L99

\bibitem[\protect\citeauthoryear{{Milgrom}}{{Milgrom}}{1983}]{1983ApJ...270..3%
65M}
{Milgrom} M.,  1983, \apj, 270, 365

\bibitem[\protect\citeauthoryear{{Rosenberg} \& {Schneider}}{{Rosenberg} \&
  {Schneider}}{2000}]{2000ApJS..130..177R}
{Rosenberg} J.~L.,  {Schneider} S.~E.,  2000, \apjs, 130, 177

\bibitem[\protect\citeauthoryear{{Rosenberg} \& {Schneider}}{{Rosenberg} \&
  {Schneider}}{2003}]{2003ApJ...585..256R}
{Rosenberg} J.~L.,  {Schneider} S.~E.,  2003, \apj, 585, 256

\bibitem[\protect\citeauthoryear{{Roychowdhury} et~al.,}{{Roychowdhury}
  et~al.}{2011}]{2011MNRAS.414L..55R}
{Roychowdhury} S.,  et~al., 2011, \mnras, 414, L55

\bibitem[\protect\citeauthoryear{{Rubin} \& {Ford} Jr.}{{Rubin} \&
  {Ford}}{1970}]{1970ApJ...159..379R}
{Rubin} V.~C.,  {Ford} Jr. W.~K.,  1970, \apj, 159, 379

\bibitem[\protect\citeauthoryear{{Safronov}}{{Safronov}}{1960}]{1960AnAp...23.%
.979S}
{Safronov} V.~S.,  1960, Annales d'Astrophysique, 23, 979

\bibitem[\protect\citeauthoryear{{Sakai} et~al.,}{{Sakai}
  et~al.}{2000}]{2000ApJ...529..698S}
{Sakai} S.,  et~al., 2000, \apj, 529, 698

\bibitem[\protect\citeauthoryear{{Sanders} \& {McGaugh}}{{Sanders} \&
  {McGaugh}}{2002}]{2002ARA&A..40..263S}
{Sanders} R.~H.,  {McGaugh} S.~S.,  2002, \araa, 40, 263

\bibitem[\protect\citeauthoryear{{Silk}}{{Silk}}{1997}]{1997ApJ...481..703S}
{Silk} J.,  1997, \apj, 481, 703

\bibitem[\protect\citeauthoryear{{Silk}}{{Silk}}{2005}]{2005MNRAS.364.1337S}
{Silk} J.,  2005, \mnras, 364, 1337

\bibitem[\protect\citeauthoryear{{Toomre}}{{Toomre}}{1964}]{1964ApJ...139.1217%
T}
{Toomre} A.,  1964, \apj, 139, 1217

\bibitem[\protect\citeauthoryear{{Tully} \& {Fisher}}{{Tully} \&
  {Fisher}}{1977}]{1977A&A....54..661T}
{Tully} R.~B.,  {Fisher} J.~R.,  1977, \aap, 54, 661

\bibitem[\protect\citeauthoryear{{Tully} \& {Fouque}}{{Tully} \&
  {Fouque}}{1985}]{1985ApJS...58...67T}
{Tully} R.~B.,  {Fouque} P.,  1985, \apjs, 58, 67

\bibitem[\protect\citeauthoryear{{Verheijen} \& {Sancisi}}{{Verheijen} \&
  {Sancisi}}{2001}]{2001A&A...370..765V}
{Verheijen} M.~A.~W.,  {Sancisi} R.,  2001, \aap, 370, 765

\bibitem[\protect\citeauthoryear{{Zwicky}}{{Zwicky}}{1937}]{1937ApJ....86..217%
Z}
{Zwicky} F.,  1937, \apj, 86, 217

\end{thebibliography}

\bsp

\label{lastpage}

\end{document}